\documentclass[10pt,conference]{IEEEtran}
\IEEEoverridecommandlockouts

\usepackage[utf8]{inputenc}
\usepackage{hhline}
\usepackage{dsfont}
\usepackage{mleftright}
\usepackage{blindtext}
\usepackage{datetime}
\usepackage{amsmath,amssymb,amsfonts}
\usepackage{algorithmic}
\usepackage{textcomp}
\usepackage{xcolor}
\usepackage[ruled,vlined]{algorithm2e}
\usepackage{mathtools}
\usepackage{graphics}
\usepackage{mathrsfs}
\usepackage{cite}
\usepackage{makecell}
\usepackage{commath}
\usepackage{multirow}  
\usepackage{graphicx}
\usepackage{comment}
\usepackage{diagbox}

\newcommand{\RNum}[1]{\uppercase\expandafter{\romannumeral #1\relax}}

\DeclareMathOperator*{\argmin}{arg\,min}
\DeclareMathOperator{\sinc}{sinc}

\def\z{{\mathbf z}}

\def\E{{\mathbb E}}
\usepackage{physics}

\begin{document}

\title{Surface Detection for Sketched Single Photon Lidar \thanks{This work was supported by the ERC Advanced grant, project C-SENSE, (ERC-ADG-2015-694888). Mike E. Davies is also supported by a Royal Society Wolfson Research Merit Award.}
}

\author{\IEEEauthorblockN{Michael P. Sheehan}
\IEEEauthorblockA{\textit{Institute of Digital Communications} \\
\textit{University of Edinburgh}\\
Edinburgh, UK \\
michael.sheehan@ed.ac.uk}
\and
\IEEEauthorblockN{Juli\'an Tachella }
\IEEEauthorblockA{\textit{Institute of Digital Communications} \\
\textit{University of Edinburgh}\\
Edinburgh, UK \\
julian.tachella@ed.ac.uk}
\and
\IEEEauthorblockN{Mike E. Davies}
\IEEEauthorblockA{\textit{Institute of Digital Communications} \\
\textit{University of Edinburgh}\\
Edinburgh, UK \\
mike.davies@ed.ac.uk}
}

\maketitle

\begin{abstract}
Single-photon lidar devices are able to collect an ever-increasing amount of time-stamped photons in small time periods due to increasingly larger arrays, generating a memory and computational bottleneck on the data processing side. Recently, a sketching technique was introduced to overcome this bottleneck which compresses the amount of information to be stored and processed.
The size of the sketch scales with the number of underlying parameters of the time delay distribution and not, fundamentally, with either the number of detected photons or the time-stamp resolution. In this paper, we propose a detection algorithm based solely on a small sketch that determines if there are surfaces or objects in the scene or not. If a surface is detected, the depth and intensity of a single object can be computed in closed-form directly from the sketch. The computational load of the proposed detection algorithm depends solely on the size of the sketch, in contrast to previous algorithms that depend at least linearly in the number of collected photons or histogram bins, paving the way for fast, accurate and memory efficient lidar estimation. Our experiments demonstrate the memory and statistical efficiency of the proposed algorithm both on synthetic and real lidar datasets.
\end{abstract}


\section{Introduction}\label{Sec: Introduction}
Single photon counting light detection and ranging (lidar) has become a prominent tool for 3D depth imaging 
due to its capacity to obtain high depth resolution measurements~\cite{manipop,lidarauto} 
using low-power, eye-safe laser sources~\cite{Pawlikowska2017SinglephotonTI}. 
The technique consists of emitting a series of light pulses and detecting each individual photon as they arrive with a single photon avalanche diode (SPAD). When an individual photon is incident on the SPAD, an avalanche of electrical charge carriers is induced that is directly detectable as a digital signal \cite{lidarauto} and can be further equated to a time delay between the emission of a light pulse and the detection of a photon by a time-correlated single photon counting (TCSPC) device. The ensemble of individual photon time delays can be characterised by a histogram, where the mode (a peak) determines the distance of an object or surface in the field of view. Multiple peaks in the  histogram can suggest the presence of multiple surfaces of varying intensity in the scene being captured. The image restoration task reduces to inferring the positions and intensities of the peaks for each pixel in the image.
\par Typically, either the time-stamp information of all the detected photons or the histogram of bin width resolution $\Delta t$ has to be recorded, stored in memory and, in general, transferred off-chip for estimation of the location and intensity parameters. The rapid emergence of high resolution, fast frame rate lidar devices~\cite{RoyoLidarReview} pose a significant challenge for both the storage and transfer of large volumes of time-of-flight information. Moreover, as the 3D data cube scales with either the resolution $\Delta t$ or the number of photons $n$, most existing 3D reconstruction algorithms in the literature~\cite{maksymova2018review} have computational complexity that also scale with these parameters. A host of methods have been proposed to tackle the memory, computational and data transfer complexities associated to high photon count, high resolution lidar data. Henderson et al.~\cite{hardware} proposed a method that uses a gated procedure to coarsely bin the time stamp of detected photons. A coarser bin length can reduce the overall size of the histogram, leading to a reduced memory requirement and smaller data transfer. In a similar vein, Zhang et al. \cite{30frames} propose a method of reducing the data transfer of photon detection times by performing a coarse to fine bin approximation of the original histogram. 
In both of these methods, a significant trade-off between compression and temporal resolution exists which could eventually ensue in a suboptimal image reconstruction without making full use of the high resolution potential of the lidar device. 
Compressed sensing based approaches \cite{4472247,7178153,comparestudy} provide an alternative compression scheme that have been proposed to exploit the sparsity of natural images in some domain to reduce signal acquisition and the overall information needed to store in memory and transfer off-chip. These methods compress along the spatial domain only and do not tackle the high resolution and/or high photon dimensions suffered in the temporal domain. 

\par A novel sketching based approach was recently proposed in \cite{sheehan2021sketchedlidar} as a solution to the data transfer bottleneck that does not suffer from an inherent trade-off between compression and temporal resolution. A compact representation, or a so-called sketch, of the time-of-flight (ToF) distribution can be computed using the detected photon time-stamps in an online manner. Fundamentally, the size of the sketch scales with the parameters of the ToF model (i.e. the positions and intensities of the objects) and is independent of both the temporal resolution $\Delta t$ and the number of photons $n$. It was shown on real-life datasets that the compression rate could amount to up to $150 \times$ whilst retaining all the salient information required to estimate the parameters of the ToF model. In this paper, we extend the sketched lidar framework by developing a surface detection algorithm
that can detect the presence of object/surface in the scene. In contrast to previous surface detection algorithms~\cite{Altmann_detection,Tachella_FastDetection,tachellaNComms} which have a complexity which is at least linear in the number of photons or histogram bins, our method only requires access to the sketch, and has a computational load proportional to the sketch size.

The proposed algorithm therefore paves the way for an extremely quick 3D depth imaging reconstruction algorithm where the majority of pixels have no object, which is often the case. Below we outline the main contributions of the paper.
\begin{itemize}
    \item We propose a detection algorithm that only requires information from a small sketch.
    \item We study the robustness of the sketched detection algorithm for different sketch sizes, signal-to-background ratios and photon counts.
    \item We analyse the algorithm on a real lidar dataset, demonstrating its good performance in the presence of high background noise.
\end{itemize}
\noindent The paper is organised as follows. In Section \ref{Sec: Sketched Lidar}, the background of sketched lidar is discussed. In Section \ref{Sec: Sketch Detection}, we propose the sketched detection algorithm. In Section \ref{Sec: Results}, we analyse the performance of the sketched detection algorithm on both synthetic and real-life datasets. We finish with a discussion in Section \ref{Sec: Discussion and Conclusion}.


\section{Sketched Lidar}\label{Sec: Sketched Lidar}
Assume there are $K$ distinct reflecting surfaces in the field of view, and denote by $\alpha_k$ and $\alpha_0$ the probability that the detected photon originated from the $k$th surface and background sources, respectively. For $1\leq p\leq n$, one can model the time of arrival of the $p$th photon detected, denoted by $x_p\in[0,T-1]$, by the following mixture distribution \cite{altmann2}
\begin{equation}
    \label{Eqn: Observ model}
    \pi(x_p|\alpha_0,...,\alpha_{K},t_1,...,t_K)= \sum^K_{k=1}\alpha_k\pi_s(x_p|t_k)+\alpha_0\pi_b(x_p),
\end{equation} 
where $\sum^K_{k=0}\alpha_k=1$. The distribution of the photons originating from the signal is defined by $\pi_s(x_p|t)=h(x_p-t)/H$, where the impulse response of the system and its associated integral are denoted by $h$ and $H=\sum_{t=1}^T h(t)$, respectively. The distribution of photons originating from background sources is in general uniformly distributed, $\pi_b(x_p)=1/T$, over the interval $[0,T-1]$.
\par In \cite{sheehan2021sketchedlidar} to tackle the data transfer bottleneck associated with high resolution, high rate single photon lidar. Central to the framework is the construction of a compact representation of the time-of-arrival data that encodes sufficient information required to infer the parameters, $\theta\coloneqq(\alpha_0,\dots,\alpha_K,t_1,\dots,t_K)$, of the observation model in (\ref{Eqn: Observ model}). The compact representation, or so-called sketch, denoted by $\z_n\in\mathbb{C}^m$ is defined by
\begin{equation}
    \label{Eqn: The sketch}
    \z_n\coloneqq\frac{1}{n}\sum^{n}_{i=1}\Phi_\omega(x_i),
\end{equation}
where $\Phi_\omega(x)=[e^{ {\rm i}\omega_jx}]_{j=1}^m$ is the feature function associated with the sketch and ${\rm i}=\sqrt{-1}$. As will be discussed further, the size of the sketch typically scales solely with the number of surfaces in the scene such that $m=\mathcal{O}(K)$. The sketch has the favourable property that it can be updated in an online fashion with each incoming photon throughout the duration of the acquisition time. Thereafter, only the resultant sketch $\z_n$ needs to be stored and/or transferred off-chip to further estimate the parameters $\theta$ of the observation model.
\par The reader may notice that the sketch is equivalent to the empirical characteristic function sampled at frequencies $\omega_j$, and $\Psi_\pi(\omega)=\E_{x\sim\pi}\Phi(x)$ is the corresponding expected characteristic function (CF) \cite{10.2307/2958763,carrasco2000generalization}. The CF has the special property that it exists for all probability distributions and captures all the information of the distribution, providing a one-to-one correspondence. For a single depth observation model ($K=1$), we define the CF of the observation model in (\ref{Eqn: Observ model}) by
\begin{equation}
\label{Eqn: Char function Obs Model}
\Psi_\pi(\omega)=\alpha_1\hat{h}(\omega)e^{{\rm i}\omega t}+\alpha_0\sinc(\omega T/2),
\end{equation}
where $\hat{h}$ denotes the (discrete) Fourier transform of the impulse response function $h$.

It is well documented in the empirical characteristic function (ECF) literature e.g. \cite{10.2307/2958763,10.2307/1912775,gemmhall}, that a sketch $\z_n$ computed over a finite dataset $\mathcal{X}=\{x_1,\dots,x_n\}$, satisfies the central limit theorem. Formally, a sketch $\z_n\in\mathbb{C}^m$ converges asymptotically to a Gaussian random variable
\begin{equation}
\label{Eqn: sketch central limit theorem}
    \z_n\xrightarrow[]{\text{dist}}\mathcal{N}\big(\Psi_\pi,n^{-1}\Sigma_\theta\big),
\end{equation}
where $\Sigma_\theta\in\mathbb{C}^{m\times m}$ has entries $(\Sigma_\theta)_{ij}=\Psi_\pi(\omega_i-\omega_j)-\Psi_\pi(\omega_i)\Psi_\pi(-\omega_j)$ for $i,j=1,2,\dots,m$. 

\par The sketched lidar inference task reduces to solving the following optimization problem
\begin{equation}
    \label{Eqn: CL loss function}
    \hat{\theta} = \argmin_\theta \lVert\z_n-\E_{x\sim\pi}\Phi(x)\rVert^2_\mathbf{W},
\end{equation}
where $\mathbf{W}\in\mathbb{C}^{m\times m}$ is a positive definite Hermitian weighting matrix chosen as the precision matrix $\mathbf{W}=\Sigma_\theta^{-1}$. The estimator is asymptotically optimal in the sense that it minimises the variance of the estimator $\hat{\theta}$ \cite{gemmhall}. 
 \par A novel sampling scheme was proposed in \cite{sheehan2021sketchedlidar} that selected frequencies $\omega$ which ensured no background photon information (noise) would be encoded in the sketch. By selecting the frequencies $\omega_j=2\pi j/T$ for $j\in[1,T-1]$ (i.e. avoiding the zero frequency of the finite basis), it can be seen that the CF $\Psi_\pi$ in (\ref{Eqn: Char function Obs Model}) is only sampled at regions where $\sinc(\omega T/2)=0$, resulting in a sketch that is effectively \textit{blind} to background noise. As a consequence, the estimates $\hat{\theta}$ that minimise (\ref{Eqn: CL loss function}) are unbiased from the presence of photons originating from background sources. Fundamentally, the size of the sketch $m$ scales with the degree of parameters in the observation model (i.e. the number of surfaces in the scene) and, crucially, is independent of both the resolution $\Delta t$ and the number of photons $n$. The sketched lidar framework therefore enables significant compression of the time-of-arrival data without sacrificing temporal resolution or estimation accuracy.

\section{Detection Algorithm}\label{Sec: Sketch Detection}
\subsection{Motivation}
If there are no objects present in the line-of-sight of the lidar device (outdoor setting), the recorded pixels 
will only consist of photon detections corresponding to background illumination, i.e. $K=0$ and $\alpha_0=1$ in \eqref{Eqn: Observ model}. Detecting and discarding pixels without peaks can avoid estimating non-existing surfaces, while reducing the computational load of posterior depth estimation. In this work, we propose a robust and fast goodness-of-fit test, which decides whether a given pixel only contains background photons. 

\subsection{Goodness-of-Fit Test}\label{subsec: Decision Rule}
The decision rule that forms the basis of the sketched detection test is centred around accepting or rejecting a hypothesis on a candidate observation model. By $H_0$, we define the null hypothesis by 
\begin{equation}
    \label{Eqn: Null hypthesis}
    H_0 : \pi(x) = \pi_b(x) 
\end{equation}
and the alternative hypothesis $H_1$ by
\begin{equation}
    \label{Eqn: Alt hypthesis}
    H_1 : \pi(x) \neq \pi_b(x)
\end{equation}

\subsection{Histogram-Based Tests}
The hypothesis test is equivalent to testing if the photons are distributed according to an homogeneous Poisson process. Under the inter-arrival time, $\Delta x = x_{i+1}-x_{i}$ is distributed according to an exponential random variable with parameter $n/T$. Hence, a standard test consists of computing the Kolmogorov-Smirnov (K-S) statistic using the empirical inter-arrival time distribution. However, this test has  important drawbacks. First, the statistic requires storing all the time-stamps, scaling linearly in the number of collected photons $n$ or histogram size $T$. Secondly, the test cannot account for the discrete nature of the time-stamps collected by the TCSPC device. 

An alternative amenable to discrete time-stamps consists in checking whether the photon count in all $T_r\leq T$ bins of a coarse histogram have a mean close to $n/T_r$ (the expected number of photons under $H_0$), using a $\chi$-squared test. In this setting, if $T_r$ is small, small peaks can be hidden in the coarse depth resolution, hindering the detection and posterior depth estimation performance. On the other hand, if $T_r$ is too large, a small number of photons per bin would depart significantly from the Gaussianity assumption of the $\chi$-squared test, degrading the performance of the method. This trade-off is shown in the experiments in Section \ref{Sec: Results}. 

\subsection{Sketch-Based Test}
The sketches described in Section~\ref{Sec: Sketched Lidar} can be applied to discrete TCSPC time-stamps, and they also provide a solution for the trade-off between depth resolution and spatial, while only requiring a very small number of statistics $m$. 

As explained in Section \ref{Sec: Sketched Lidar}, the sketches converge quickly to a Gaussian distribution, and hence they are amenable to use in a $\chi$-squared test.
This test has been used extensively throughout the ECF literature to form hypothesis, normality and detection tests based on the characteristic function and its empirical counterpart \cite{Fan_ECFTest,koutrouvelis1981goodness}. Under the assumption of no surfaces, we have $\mathbb{E}  \{z_n\} = 0$ and $\Sigma_\theta$ is just the identity, hence
\begin{equation}
    \label{eqn: Mah Distance Test statistic}
    D^2\coloneqq \| \z_n\|_2^2,
\end{equation}
is used as the test statistic to form a one-sided hypothesis test for $H_0$ against $H_1$. 
Under the null hypothesis $H_0$, the squared distance follows $D^2\xrightarrow[]{\text{dist}}\chi^2_\nu$ in distribution with $\nu=m-2K-1$ degrees of freedom \cite{koutrouvelis1981goodness}. One can therefore reject the null hypothesis $H_0$ at significance level $\beta$ if $D^2>\bar{z}_\beta$ where $\bar{z}_\beta$ is the upper $\beta$-percentile of the $\chi^2_\nu$ distribution. 

\par Importantly, the decision rule is based solely on the sketch of size $m$. This is significant as (i) the full data of the TCSPC histogram is not required in the computation and can be discarded from memory (ii) the squared test statistic can be computed in $\mathcal{O}(m)$. 

\subsection{Extension to Non-Constant Background}
In some practical settings, the distribution of background photons $\pi_b$ might not be exactly constant, for example due to pile-up phenomena~\cite{rapp2021high}. In these cases, the sketched test statistic can be easily modified to account for a data-driven $\hat{\pi}_b$, using background photons collected in a calibration step, 
\begin{equation}
    \hat{\z}_n = \mathbb{E}_{\hat{\pi}_b} \{\Phi_\omega(x_i) \}.
\end{equation}
The test statistic is then $D^2\coloneqq \| \z_n - \hat{\z}_n\|_2^2$. It is worth noting that the data-driven test can be interpreted as a random features version of the maximum mean discrepancy~\cite{gretton2012kernel}.
 

\subsection{Spatial Regularization}\label{subsec: TV Reg}
Neighbouring pixels in a lidar scene typically exhibit the same number of surfaces owing to spatial correlation. Exploiting the inherent spatial correlation in typical lidar scenes can further reduce the occurrence of false positives. In \cite{Tachella_FastDetection}, Tachella et al. proposed a total variation (TV) based spatial regularization that created a more homogeneous map of the present targets. Here we include a similar spatial regularization based on the goodness-of-fit. Formally, the TV based spatial regularization is defined by the map
\begin{equation}
  \hat{\vb*{v}}\coloneqq\mathcal{H}_{0/1}\left(\argmin_{\vb*{v}}\lVert\vb*{v}-\vb*{y}\rVert^2_2+\tau\lVert\vb*{v}\rVert_{\text{TV}}\right)  
\end{equation}
where the input image $\vb*{y}$ contains the $\chi$-squared statistic $D^2$ of pixel $(i,j)$, $\lVert\cdot\rVert_{\text{TV}}$ is the isotropic TV operator, $\tau$ is a user-defined regularization parameter and $\mathcal{H}_{1/0}$ is a hard-thresholding operator which assigns 1 to positive inputs and 0 otherwise. In Section \ref{Sec: Results}, we show this spatial regularization can help remove a proportion of false positive alarms producing a more homogeneous detection map. 

\subsection{Posterior Depth Estimation}
Once the pixels containing at least one surface have been identified, the depth is estimated by solving the optimization problem in \eqref{Eqn: CL loss function}, as detailed in  \cite{sheehan2021sketchedlidar}. The depth estimation also scales only in the number of sketches $m$.



\section{Empirical Results}\label{Sec: Results}
\begin{figure}[ht!]
\centering
\includegraphics[scale=0.25]{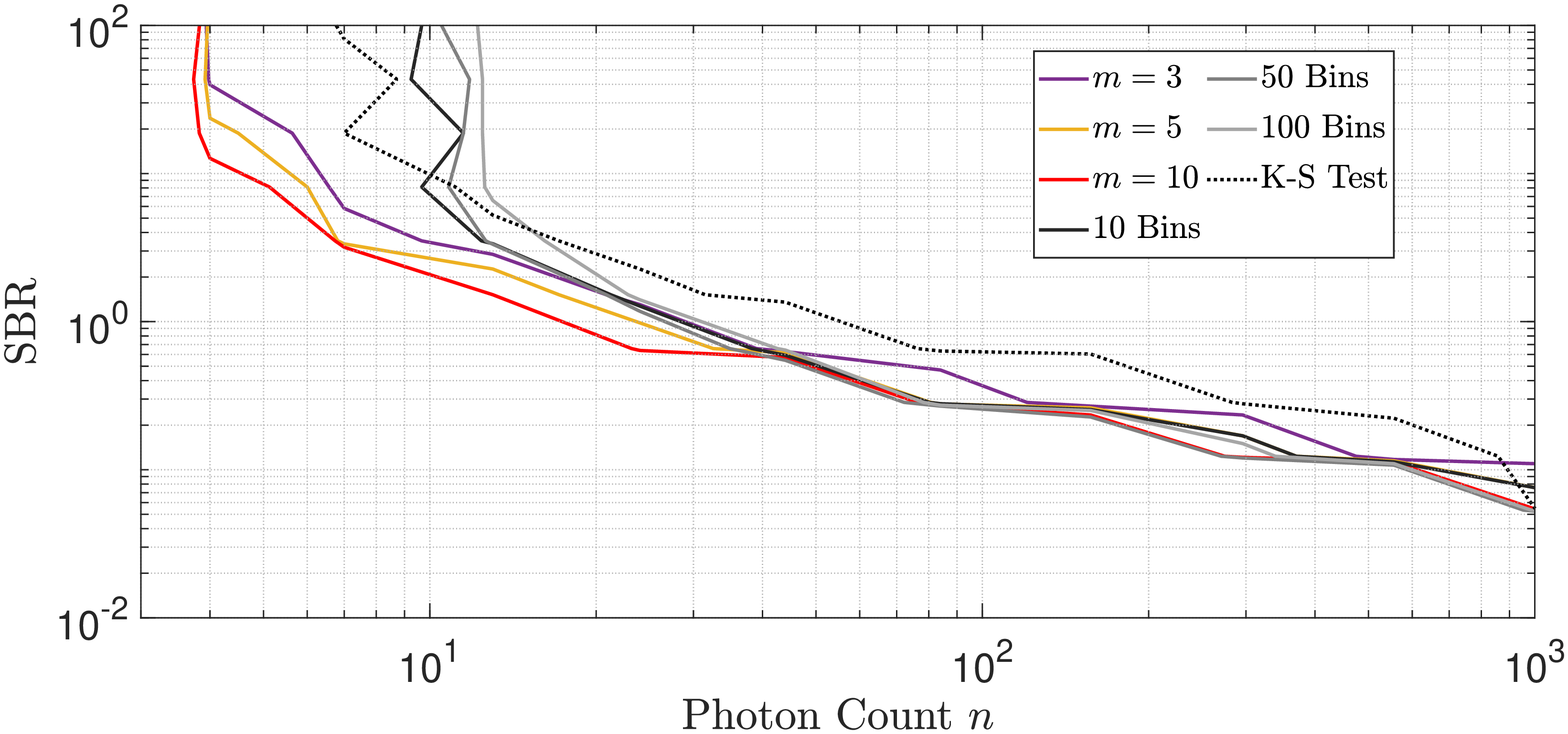}
\caption{Detection performance of the sketch based method for sketch sizes of $m=3,5,10$, the coarse histogram test for histograms of size $T_r=10,50,100$, and the full data K-S test. The graphs correspond to a detection probability of 95\%.}
\label{fig: PD compare}
\end{figure}

\begin{figure}[ht!]
\centering
\includegraphics[scale=0.25]{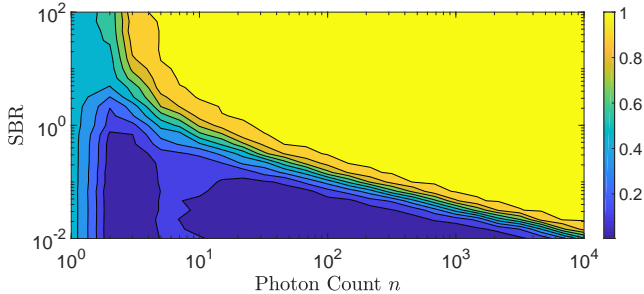}
\caption{Empirical probability of detection for the proposed sketch-based detection scheme using a sketch size $m=10$.}
\label{fig: MD PD}
\end{figure}

\begin{figure}[ht!]
\centering
\includegraphics[scale=0.25]{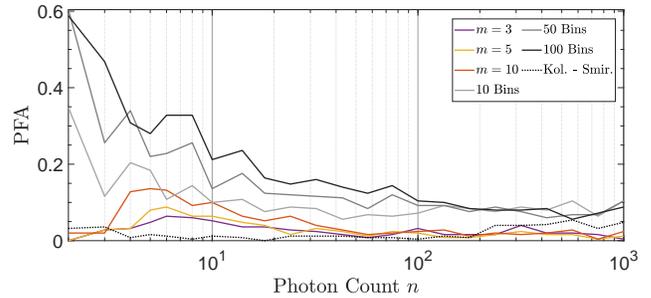}
\caption{Probability of false alarm of the sketch based method for sketch sizes of $m=3,5,10$, the coarse histogram test for histograms of size $T_r=10,50,100$, and the full data K-S test. }
\label{fig: PFA compare}
\end{figure}

In this section, we evaluate the sketch-based detection scheme on both real and synthetic data. 
First, we analyse the effect the signal-to-background ratio (SBR), defined by SBR$=\alpha/(1-\alpha)$, and the photon count $n$ on both the true positive and false alarm rate, using a Gaussian impulse response with standard deviation $\sigma=T/100$, for $T=5000$. Figure \ref{fig: MD PD} shows a map of the empirical probability of detecting a single peak for various SBR's and photon counts for the proposed sketch-based detection. Even for moderately high SBR, for example SBR$=1$, the detection scheme only requires approximately 20 photons to achieve detecting the single peak with high probability.  Figure \ref{fig: PD compare} shows the SBR/photon count level-curves for a true positive rate of $95\%$ for various sketch sizes and full-data approaches. For the full-data approach, a $\chi^2$ test was constructed on the true observation model in (\ref{Eqn: Observ model}) where adjacent bins were concatenated to maximise the power of the hypothesis test. For reference, we also include a K-S test  which is again performed on the full data (see \cite{KS_PoissTest} for details). For each test the significance level was set at $\beta=0.05$. Figure \ref{fig: PFA compare} depicts the empirical probability of false alarm (PFA) as a function of the photon count for various sketch sizes and for the aforementioned full data hypothesis tests. 

\par
\begin{figure}[ht!]
\centering
\includegraphics[scale=0.26]{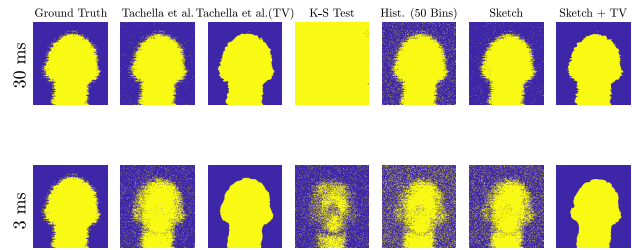}
\caption{Detection maps of the polystyrene dataset \cite{Altmann_detection} for the proposed sketch and sketch plus TV detection schemes in comparison with other non-compression detection techniques. }
\label{fig: MD face compare}
\end{figure}

\begin{figure}[ht!]
\centering
\includegraphics[scale=0.26]{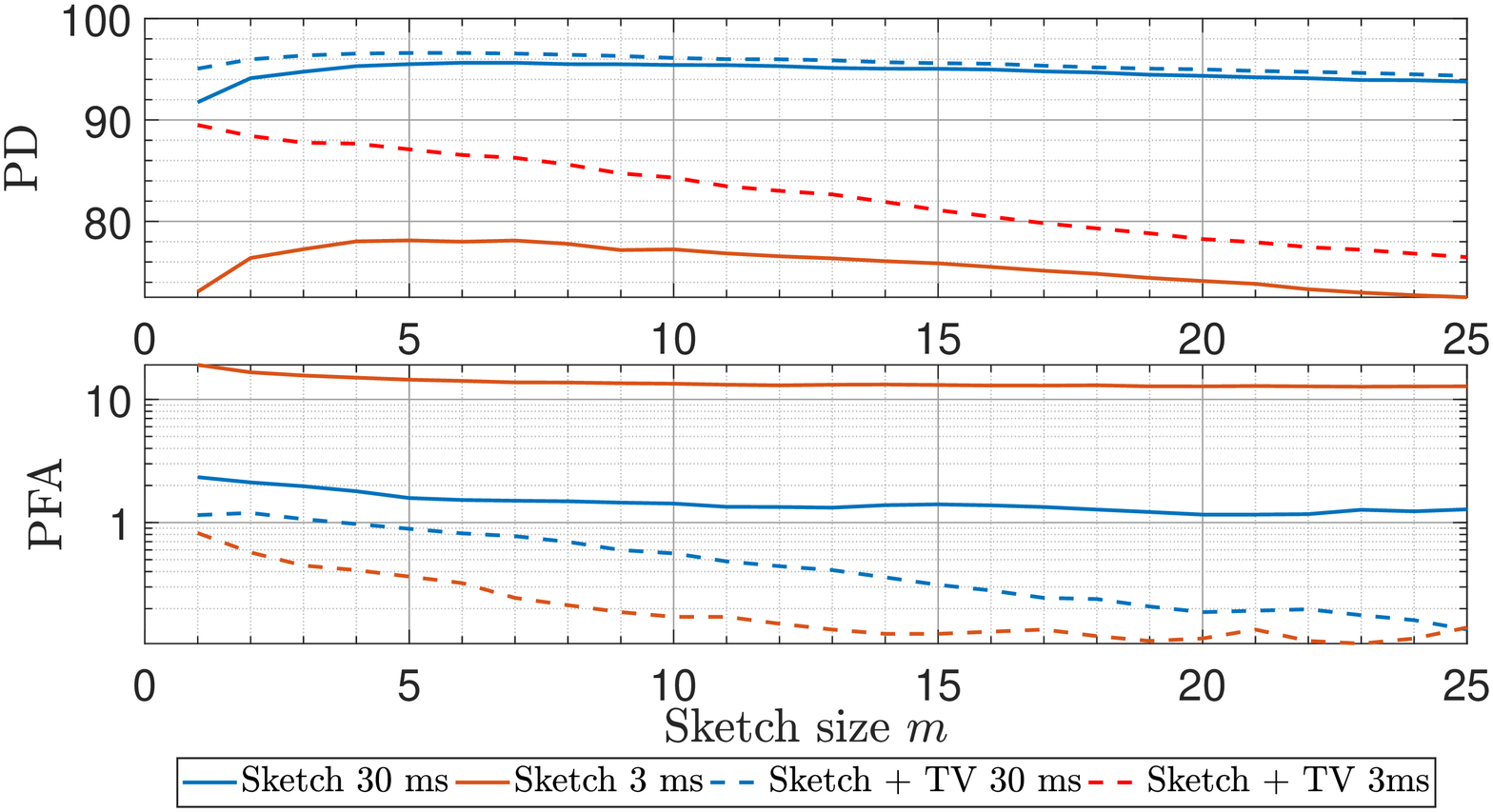}
\caption{Empirical probabilities of detection (top) and false alarm (bottom) for the evaluated detection methods using the polystyrene head dataset.}
\label{fig: Mah. dist. face compare}
\end{figure}

 Next, we compare the proposed sketch-based detection algorithm with the $\chi^2$ test on the full data observation model as well as the two detection methods proposed by Tachella et al. in \cite{Tachella_FastDetection}, using a real lidar dataset consisting of a polystyrene head measured at a stand-off distance of 325 metres. See details of the dataset in \cite{Altmann_detection}. The dataset consists of $200\times 200$  pixels with $T=2700$ histogram bins per pixel and an approximate SBR of 0.29. Figure \ref{fig: MD face compare} shows the detection maps for two different per-pixel acquisition times (30 ms and 3 ms) corresponding to an average photon count of 900 and 90 photons, respectively. The sketch size was set at $m=5$ and the significance level was set at 0.05 and 0.2 for the 30 ms and 3 ms acquisition times, respectively. Also included is the proposed sketch-based method with spatial TV regularization as discussed in Section \ref{subsec: TV Reg}.  The PD and the PFA for each acquisition time are shown in Table \RNum{1} for each detection scheme. The PD and PFA for both the sketch and sketch plus TV regularization are depicted in Figure \ref{fig: Mah. dist. face compare} for increasing sketch size $m$. 

\begin{table}[ht!]
\centering
\begin{tabular}{c|c|c|c|c|}
\cline{2-5}
 & \multicolumn{2}{l|}{PD $\%$} & \multicolumn{2}{l|}{PFA $\%$} \\ \cline{2-5} 
& 30ms          & 3ms          & 30ms          & 3ms           \\ \hline
\multicolumn{1}{|l|}{Tachella et al.}         & 98.5          & 82.3         & 0.7           & 6.5           \\ \hline
\multicolumn{1}{|l|}{Tachella et al. (TV)}    & 98.4          & 93.7         & 3.5           & 1.1           \\ \hline
\multicolumn{1}{|l|}{K-S Test (Full Data)} & 100          & 49.9         & 99.3         & 6.9         \\ \hline
\multicolumn{1}{|l|}{Hist. (50 Bins)} & 97.5          & 77.6         & 8.8           & 19.9          \\ \hline
\multicolumn{1}{|l|}{Sketch}                  & 95.4          & 77.2         & 1.4           & 14.4          \\ \hline
\multicolumn{1}{|l|}{Sketch + TV}             & 96.6          & 88.1         & 0.9           & 0.5           \\ \hline
\end{tabular}
\label{Table: Mah. dist. Face Compare}
\caption{Probabilities of detection (PD) and probabilities of false alarm (PFA) for the proposed sketch-based detection schemes and other detection algorithms. The sketch size is set at $m=5$ and the full data $\chi^2$ test was chosen using 50 adjacent bins to optimise the PD/PFA trade-off.}
\end{table}

\par The results show that on both synthetic and real datasets the sketch-based detection scheme achieves a similar, or better, PD/PFA trade-off than the full data $\chi^2$ detection test. In fact the sketch-based detection scheme achieves a far lower PFA than the full data $\chi^2$ detection for both acquisition times. 
\

\section{Conclusion}\label{Sec: Discussion and Conclusion}
In this paper, we developed a detection scheme based solely on a compact representation sketch that is robust in detecting the presence of a surface for each pixel in the lidar scene. As a result, pixels consisting of non-existing surfaces can be discarded from memory reducing the overall computational and memory load of transferring and reconstructing a lidar scene. Moreover, it is shown that only a minimal sized sketched is required to achieve a high probability of detection on both synthetic and real datasets, achieving a better PD/PFA trade-off than the corresponding $\chi$-squared test on the original histogram data. The proposed sketch-based detection algorithm paves the way for high accuracy 3D imaging at fast frame rates with low power consumption without sacrificing the overall temporal resolution. In Section \ref{Sec: Results}, we considered detecting the presence of a single surface. However, the framework can be readily used for distinguishing between single and multiple surfaces. As a result, further computational and memory savings can be achieved, although we leave further analysis and experiments of multi-surface detection for future work. 

\bibliographystyle{IEEEtran}
\footnotesize
\bibliography{ref.bib} 
\newpage
\end{document}